\documentclass{article}
\usepackage{graphicx} 
\usepackage[hyphens]{url} 
\usepackage{natbib}
\usepackage{url}
\usepackage[breaklinks]{hyperref} 
\usepackage{adjustbox} 
\usepackage{float}
\usepackage{placeins}
\usepackage{indentfirst}
\usepackage[hyphens]{url}   
\usepackage{hyperref}       
\usepackage{lipsum} 
\usepackage{array}  
\usepackage{tabularx}
\usepackage{longtable}
\usepackage{booktabs}

\DeclareUnicodeCharacter{2060}{}

\title{Comparative Global AI Regulation: Policy Perspectives from the EU, China, and the US}
\author{
    Jon Chun\textsuperscript{1} \\
    Christian Schroeder de Witt\textsuperscript{2} \\
    Katherine Elkins\textsuperscript{1} \\
    \\
    \textsuperscript{1}Kenyon College \\
    \textsuperscript{2}Oxford University
}
\date{October 2024}

\begin{document}

\maketitle

\section{Abstract}

As a powerful and rapidly advancing dual-use technology, AI offers both immense benefits and worrisome risks. In response, governing bodies around the world are developing a range of regulatory AI laws and policies. This paper compares three distinct approaches taken by the EU, China and the US. Within the US, we explore AI regulation at both the federal and state level, with a focus on California's pending Senate Bill 1047. Each regulatory system reflects distinct cultural, political and economic perspectives. Each also highlights differing regional perspectives on regulatory risk-benefit tradeoffs, with divergent judgments on the balance between safety versus innovation and cooperation versus competition. Finally, differences between regulatory frameworks reflect contrastive stances in regards to trust in centralized authority versus trust in a more decentralized free market of self-interested stakeholders. Taken together, these varied approaches to AI innovation and regulation influence each other, the broader international community, and the future of AI regulation.

\section{Introduction}

Proposed in April 2021 and agreed upon by December, the EU Act was the first major coordinated effort to regulate AI; it came into force in August 2024. The Biden administration published its Blueprint for an AI Bill of Rights in October 2022. It spoke to the need to protect citizen's privacy and freedom from algorithmic discrimination. The following month, on November 25th, several Chinese government ministries jointly released regulations on AI-generated deepfakes. The US Whitehouse Executive Order \#14110 "Safe, Secure, and Trustworthy Development and Use of Artificial Intelligence" was issued in October 2023. In February 2024, California State Senator Scott Wiener introduced what was arguably the strictest AI regulation with the "SB-1047 Safe and Secure Innovation for Frontier Artificial Intelligence Models Act". After ten major revisions across both the California Senate and Assembly, the bill passed on September 3rd and now faces an uncertain future as it awaits Governor Newson's signature to become law.

Underlying these varied global efforts are common concerns over the expected social, economic, and geopolitical impacts of AI. The European Union has taken a proactive stance with regard to social effects of AI, implementing stringent regulations aimed at fostering competitiveness while prioritizing ethical considerations, enhancing privacy protections, and mitigating potential harms. China has focused on aligning AI development with "core socialist values" while also addressing issues of transparency and workers' rights. The United States, meanwhile, has grappled with concerns about AI-generated disinformation, election integrity, and the role of content recommendation systems in social media, particularly in light of events such as the 2020 elections and the rise of platforms like TikTok~\citep{smit2022}. 

Beyond the societal effects, there is universal acknowledgement of the centrality of AI to related technologies like advanced chips, energy production and storage, 5G/6G telecommunications, satellites, and robotics. These illustrate the importance of AI to national economics, competitiveness, and security. Some argue that the EU's stricter AI regulation may inadvertently stifle innovation, deter investment, and weaken Europe's position in the global AI and technology race~\citep{suominen2020}. The tension between regulatory safety and competitiveness is particularly evident in the dynamics between China and the US, with both nations striving to balance innovation with responsible AI development~\citep{itif2024chinaai}.

Still, there are efforts to transcend national policies and competitions to develop a global harmonization of AI regulation. AI automation threatens widespread job displacement, exacerbates inequality and accelerates the need to transition to a rapidly-evolving job market
\citep{Whitehouse2024c}. In September 2024, the United Nations' AI Adivsory Body released a report highlighting key areas of concern that transcend national boundaries. These include pervasive latent biases, emergence of surveillance states, and AI generated disinformation~\citep{UNAI2024}. Additionally, serious legal, security, and humanitarian issues--particularly related to autonomous weapons and public security--underscore the importance of international cooperation. As AI continues to evolve, the global community faces the complex task of developing regulatory frameworks that can effectively address these multifaceted challenges while fostering innovation and ensuring equitable benefits across nations.

\section{AI Governance}

This section describes the emerging AI regulatory frameworks in the EU, China, and the US at both federal and state levels.  In particular, it contrasts the top-down, risk-based approach of the EU AI Act with the more market-driven approach of the US. The latter emphasizes coordinating existing legal, regulatory, and enforcement entities from the federal level down to states and cities. In between is the Chinese approach, which has the appearance of centralized regulatory control, but in practice emphasizes decentralized innovation, regional competition, and economic development at the local levels.

While the EU and China appear to have relatively stable AI regulatory frameworks, there is a growing debate in the US about the future direction of AI regulation. The Biden Executive Order \#14110 on "Safe, Secure, and Transparent Development and Use of AI" coordinates over 100 specific tasks both within and between over 50 federal entities in a decentralized way that largely augments existing regulatory laws and agencies. However a number of US Congressional committees, proposals, and influential public/corporate interest groups are lobbying for a new AI regulatory structure that is more centralized, restrictive, and punitive than EO \#14110. Some even promote centralized registration of models, proofs of AI safety, and criminal penalties~\citep{LegiScan2024, schumer2023}.

\section{The European Union}

\subsection{Overview}

The 2024 EU AI Act is positioned as the world's first comprehensive AI law~\citep{eu_ai_act_2024}. Just as prior European general purpose legislation, such as the 2016 General Data Protection Regulation (GDPR)~\citep{gdpr_article7}, the EU AI Act represents complex joint efforts and interests across various EU bodies, including the European Commission, the European Parliament and the European Council, where the latter represents the Heads of State of all EU member countries. Influence on the Act's formation was also taken publicly by national government officials, such as France's premier Macron's overt lobbying for exemptions for open source AI providers such as Mistral~\citep{abboud_eus_2023}, as well as, both publicly and covertly, by lobbying and industry groups, including Big Tech~\citep{perrigo_exclusive_2023}, and German pro open-source non-profit LAION~\citep{noauthor_call_nodate}.

Uniquely, the Act was first constructed within a product safety framework, but then blended with a fundamental rights agenda at the behest of the European Parliament and against pressure by the European Commission~\citep{caroli_podcast}. This approach resulted in a unique and novel blend of legislative frameworks, clearly setting the EU AI Act apart from prior legislation building on established frameworks such as the GDPR. As Dragos Tudorache, a member of the European Parliament from Romania and the chair of the Special Committee on Artificial Intelligence in a Digital Age, remarked: "Regulation isn't just rules, it's an opportunity to express our values~\citep{tyrangiel_opinion_2024}."

While constituting an innovative syncretism at heart, the Act does follow and respect earlier European generalist regulatory initiatives, such as the GDPR, and the 2012 Digital Markets Act~\citep[DMA]{dma}. In fact, the EU started examining the compatibility of the GDPR and AI as early as in 2020~\citep{eprs2020gdpr}. Nevertheless, the release of ChatGPT in November 2022 and its rapid adoption by millions of consumers worldwide caught European policymakers off-guard and led to significant adjustments to the Act's handling of AI governance.  On December 9th, 2023, the Act was provisionally agreed between the European Parliament and the European Council. During the 3-day round of negotiations, the Act's scope was tightened~\citep{compromise}. It was clarified that the Act does not apply outside of European law, does not infringe on the security competences of the member states or so-entrusted entities, nor any military or defense applications. Importantly, it was clarified that the Act would not apply to sole purposes of research and innovation, nor other non-professional use. Most importantly, the Act's traditional approach to AI systems risk classification was complemented by the notion of general-purpose AI systems (GPAI)~\citep[Article 3(63)]{eu_ai_act_2024}, resulting in a parallel governance track for such AI systems. Despite open questions, the Act's ambition to distinguish between GPAI and non-GPAI systems, and GPAI systems of \textit{systemic risk}, is similarly unique among AI regulations globally.

\subsection{The Geopolitics of the Act}

Besides regulating the EU single market, the Act is widely regarded as a strategic effort by the European Commission to establish themselves as the leading AI rulemakers globally~\citep{noauthor_eus_2024}. Just as after the adoption of the GDPR in 2016~\citep{gdpr_article7}, it is speculated that companies across the world will begin to prioritize compliance with European AI law out of economic necessity, not through coercion~\citep{almada_brussels_2024}. In the case of the GDPR, this form of ``Brussels Effect''~\citep{bradford_brussels_2020} was complemented by a ``de jure'' effect in which countries with a lack of their own regulatory capacity, such as many developing countries,  incorporated EU laws instead. For example, the Philippines incorporated the right to be forgotten into their Data Privacy Act of 2012~\citep{philippines2012dataprivacy,linklaters}. In a similar way, it is speculated that the EU AI Act may similarly become the de-facto standard for AI governance in much of the Western and developing world~\citep{almada_brussels_2024}. 

One direct institutional consequence of the EU AI Act is the establishment of a novel authority under the helm of the European Commission, namely the EU AI Office. This new office will not only oversee AI regulation and AI systems' compliance, provide a central pool of AI expertise to the EU’s member states, but also provides a ``strategic, coherent, and effective European approach on AI at the international level''~\citep{eu_ai_office}. A specialist position, The Advisor for International Affairs, will represent the AI Office in ``global conversations on convergence toward common approaches''~\citep{eu_ai_roles}. The EU AI Office therefore is not only meant to consolidate the EU’s approach to AI regulation within the European market, but will likely centrally support the EU’s foreign policy ambitions in economic and trade negotiations.

\subsection{Laws and Regulation}

The Act is designed as an \textit{adaptive legislation}, meaning that many details are intentionally left vague to permit later adaptation as technology changes. At its core lies a risk classification system that puts obligations mostly on the developers (``providers'') of AI systems. Importantly, the Act not only applies to systems that are placed on the market or put into service in the EU, but also to AI systems whose output is used in the EU. The use of AI systems for national security, research, or recreational purposes, as well as, more generally, end-users of AI systems are excluded from regulation under the Act.

\begin{figure}
    \centering
    \includegraphics[width=\linewidth]{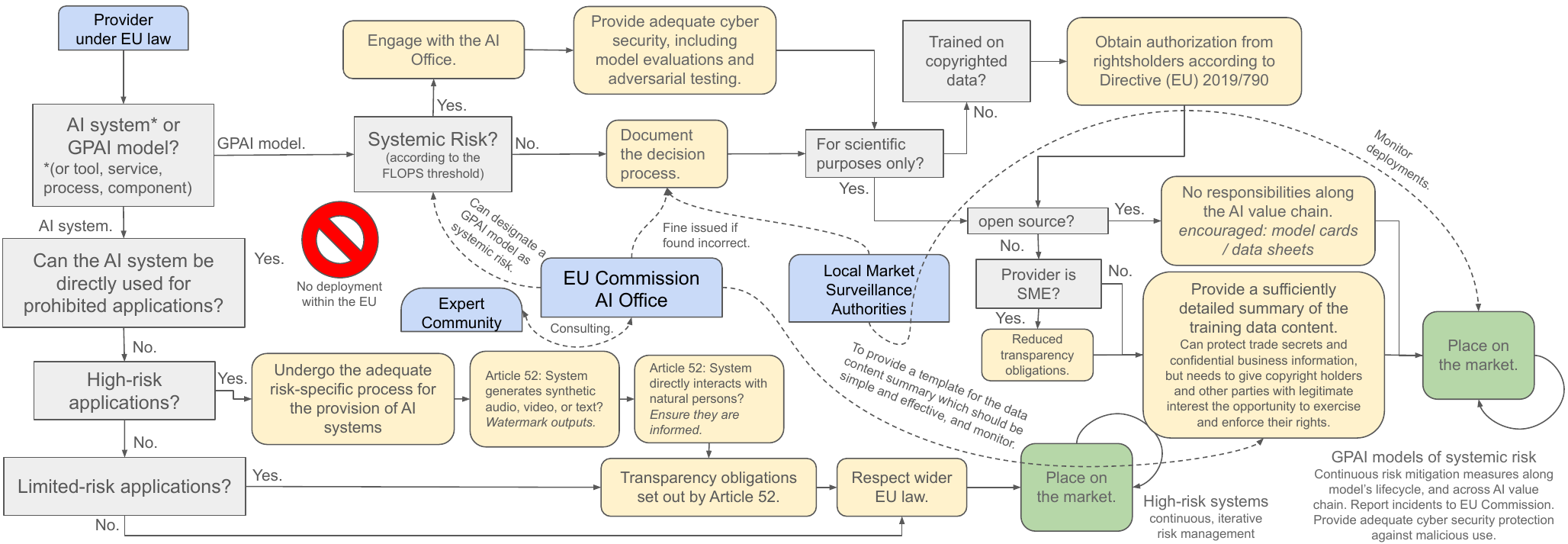}
    \caption{Decision tree for providers of (GP)AI systems and GPAI models on the way to the EU market. }
    \label{fig:figure1}
\end{figure}

\subsubsection{Systems and Models}

The AI Act adopts the definition of \textit{AI system} from the OECD's AI Principles, defining an AI system as \textit{``a machine-based system that is designed to operate with varying levels of autonomy and that can, for explicit or implicit objectives, generate outputs such as predictions, recommendations, or decisions that influence physical or virtual environments''}~\citep{oecd_ai_principles}. \textit{AI models}, conversely, are mathematical algorithms or trained models that, in isolation, lack additional components such as a user interface or hardware integration that would allow them to be used as an AI system. 

\sloppy
Among models, the Act further disambiguates between non-GPAI and \textit{general-purpose AI (GPAI) models}, with the latter defined as \textit{``an AI model, including where such an AI model is trained with a large amount of data using self-supervision at scale, that displays significant generality and is capable of competently performing a wide range of distinct tasks regardless of the way the model is placed on the market and that can be integrated into a variety of downstream systems or applications, except AI models that are used for research, development or prototyping activities before they are placed on the market''}~\citep[Article 3(63)]{eu_ai_act_2024}. Where a GPAI model is integrated into an AI system, the system is referred to as a \textit{GPAI system} provided that the system has the capability to serve a variety of purposes.

\subsubsection{Roles}

The Act fundamentally distinguishes between the roles of \textit{provider}, \textit{deployer}, \textit{importer} and \textit{distributor}. Providers are those \textit{``placing on the market or putting into service AI systems or placing on the market general-purpose AI models in the Union''}~\citep[Article 2]{eu_ai_act_2024}, irrespectively of their location. They have pre-market obligations including an initial risk assessment, risk-specific compliance, as well as risk-specific post-market obligations. Deployers are users of AI systems that are based on the EU and that don't fall within a small number of non-professional use cases. The Act furthermore distinguishes between AI models or systems provided or deployed from within the EU, and those from outside.

\subsection{Risks}

At the core of the EU AI Act lies a risk classification system for AI systems based on their possible ``direct'' use cases. In the case of GPAI systems, the degree of ``directness'' is understood as the extent to which implemented safety measures can prevent risk-relevant use by deployers. If local market surveillance believe that a GPAI systems can be (or become) ``directly'' used for high-risk activities the EU AI Office will carry out the corresponding compliance procedures~\citep{eu_ai_act_2024}. 

\paragraph{Prohibited use cases.} Integrating fundamental human rights into the product safety framework, the Act defines a class of \textit{prohibited AI practices}, such as placing on the market AI systems that can be used for certain forms of manipulation and exploitation, social scoring purposes, and certain biometric identification purposes. Furthermore, the deployment of AI systems that leave the user uninformed about their interaction with an AI system, emotion recognition systems or biometric categorisation systems, or AI systems producing deepfakes are all likewise prohibited~\citep[Article 5]{eu_ai_act_2024}.

\paragraph{High-risk use.} The class of \textit{high risk} systems constitutes the majority of risk-related regulation~\citep[Article 6]{eu_ai_act_2024}. High-risk systems include a wide variety of systems as defined in~\citep[Annex I-III]{eu_ai_act_2024}, including systems meant to serve as safety systems for other AI systems. According to~\citep[Article 28(2a)]{eu_ai_act_2024}, providers of high risk systems are subject to compliance obligations, including the establishment of risk and quality management systems, data governance, human oversight, cybersecurity measures, postmarket monitoring, and maintenance of the required technical documentation. Owing to the Act's adaptive nature, it is expected that these obligations will be further detailed in later, sector-specific regulation. 

\sloppy
\paragraph{Limited risk.} Chatbots or AI systems that generate content or aid in decision-making without any critical safety aspects or significance are deemed of \textit{limited risk}, although the Act only indirectly defines this class~\citep[Recital 32a]{eu_ai_act_2024}. These systems are merely subject to transparency obligations, including end-users of such systems must be informed that they are interacting with AI. 

\paragraph{Minimal risk.}
AI systems that pose little to no risk to users' rights, health, or safety are left unregulated by the Act, although other obligations under EU law still apply. These systems are sometimes referred to as \textit{minimal risk} although this term is not used in the Act.

\begin{figure}
    \centering
    \includegraphics[width=0.95\linewidth]{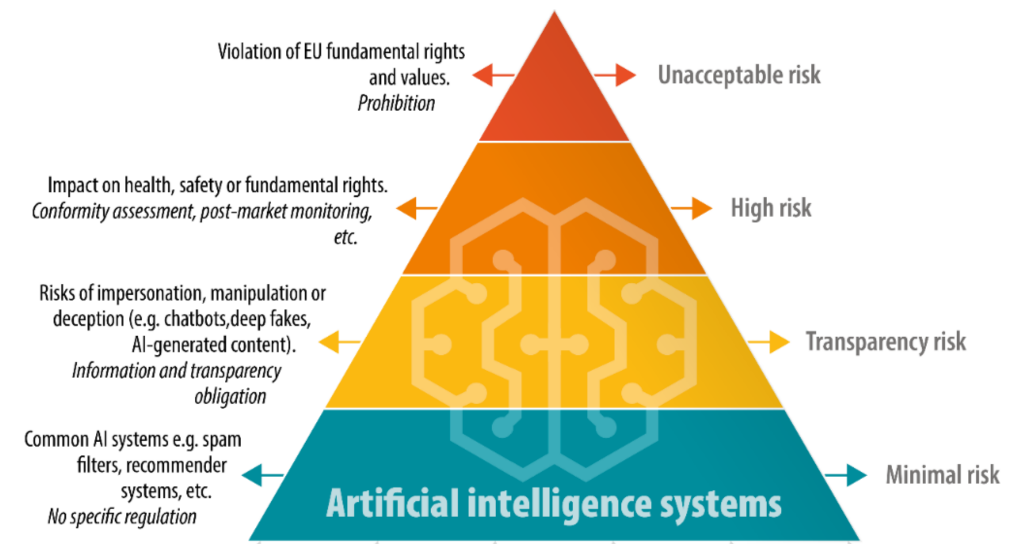}
    \caption{The risk pyramid for AI systems (taken from~\citep{madiega2024artificial})}
    \label{fig:figure2}
\end{figure}

Importantly, in the case of GPAI models, a special risk category is defined for the standalone model even before having been integrated into an AI system.

\paragraph{GPAI models of systemic risk.}
The Act imposes particular regulation on providers of general-purpose AI models of systemic risk~\citep[Article D]{eu_ai_act_2024}, which it defines to be all models for which \textit{``the cumulative amount of compute used for its training measured in floating point operations (FLOPs) is greater than $10^{25}$''}~\citep[]{eu_ai_act_2024}. The limit of $10^{25}$ was reportedly reached as a middle ground between $10^{24}$ and $10^{26}$ demanded by two opposing factions, the European Parliament, and the European Commission~\citep[Chapter II(8)]{eu_ai_act_2024}\citep{caroli_podcast}. Providers of GPAI need to register their model with the European Commission, and need to adhere to a wide-ranging catalogue of safety and security criteria. Owing to its adaptive nature, the Act purposefully leaves various technical criteria related to systemic risk classifications open for later adjustment to account for the unpredictability of technological progress. 

\subsection{Innovation and Open Source}

The Act contains several measures intended to harness the economic and societal benefits of open-source AI software~\citep{eiras_near_2024,eiras_risks_2024}. The Act defines free and open-source AI components to cover ``the software and data, including models and general-purpose AI models, tools, services or processes of an AI system'' and explicitly states that provision of such models on open repositories should not seen as a form of monetization~\citep[Recital 103]{eu_ai_act_2024}. 

\sloppy
The EU Act contains wide-ranging exemptions for providers of certain AI systems provided under free and open source software licenses~\citep[Article 53-54]{eu_ai_act_2024}. To be exempt, the systems may not contain GPAI models that fall within the systemic risk category, or otherwise exhibit unacceptable behavior. The Act distinguishes the above from providers of pre-trained AI models that are made accessible to the public under a license that allows for the access, usage, modification, and distribution of the model, and whose parameters, including the weights, the information on the model architecture, and the information on model usage, are made publicly available. It is to be noted that the term open source model is not used explicitly and the degree of legal overlap with open source software is not immediately clear, and hence such models might be referred to rather as open models. Open models are not exempt from Article \citep[C(1)(c)-(d)]{eu_ai_act_2024}, as well as \citep[Article D]{eu_ai_act_2024} and \citep[Article 28(2a)]{eu_ai_act_2024}, the latter governing third-party obligations ``along the AI value chain of providers, distributors, importers, deployers or other third parties'' ~\citep{eprs2020gdpr} for high-risk AI systems.

Under any circumstances, providers of open GPAI model providers are responsible for transparency obligations according to~\citep[Article C(1)(c)-(d)]{eu_ai_act_2024}. These transparency requirements include respecting existing Union copyright law~\citep[Article C(1)(c)]{eu_ai_act_2024} according to Article 4(3) of the Digital Single Market Directive (EU) 2019/790~\citep{dsm_directive_2019}, and the need to make a \textit{``sufficiently detailed summary of the content used for training of the general-purpose AI model''}~\citep[Article C(1)(d)]{eu_ai_act_2024}. The consequences of these transparency requirements for open GPAI model providers have been examined in~\citep{warso2024AIAct}. Importantly, Directive (EU) 2019/790 expands GDPR regulation to copyright owners who share content online, meaning that key GDPR rights, such as the right to opt out~\cite[GDPR Article 7]{gdpr_article7} would require that copyright owners could ask for their data to be removed from open GPAI model training data. 

As a further measure to stimulate innovation, member states can establish a regulatory sandbox, i.e. a controlled environment that facilitates the development, testing and validation of innovative AI systems (for a
limited period of time) before they are put on the market~\citep{madiega2024artificial}. 

\section{China}

\subsection{Overview}

China’s approach to AI governance and regulation is a hybrid between the centralized, top-down approach of the EU and the decentralized, free-market of competing interests in the US. Like the EU, China emphasizes safety, individual protections, and social harmony through top-down guidance, regulation, and enforcement~\citep{zhang2022}. Like the US, China also emphasizes bottom-up innovation and economic development through a mix of decentralized provincial control alongside very competitive local markets. This hybrid approach seeks to optimize the benefits from both the EU and US models. The EU AI Act takes a coherent, universal risk-based approach, but the abstract and ambiguous language belies the hard work of grappling with real-world details in applying these general rules to disparate, complex and highly situational cases. Conversely, a fragmented, sector-specific approach like the US EO \#14110 lacks coherent high-level simplicity, but benefits from experienced domain experts translating goals into more clear, immediate, and effective enforcement. China seeks to benefit from the coherence of the EU AI Act and the practical benefits of the US approach, which promotes innovation and economic competitiveness. 

\subsection{Laws and Regulations}

\FloatBarrier

China has advanced some of the first AI laws and regulations at the national level, which are summarized in Table 2. Unlike the horizontal risk-based approach of the EU, China has favored the sector-specific US approach of laws tailored to specific use-cases. These specific use-cases range from data privacy (November 2021) to recommendation algorithms (March 2022) to generative AI (January \& August 2023). Despite appearances of centralized government control, Chinese AI regulations are the product of an iterative process involving diverse stakeholders that include mid-level bureaucrats, academics, corporations, startups, and think tanks~\citep{sheehan2024}. The central government relies upon a pipeline of these experts to formulate, clarify, and interpret the details, while local officials mainly concern themselves with ensuring goals and outcomes are aligned with Chinese and socialist ideology~\citep{zhang2022}. Both China’s State Council and the Chinese Academy of Social Sciences have announced intentions of working towards a more holistic National AI law, although the outcome is uncertain~\citep{webster2023}.

\begin{figure}
    \centering
    \includegraphics[width=0.88\linewidth]{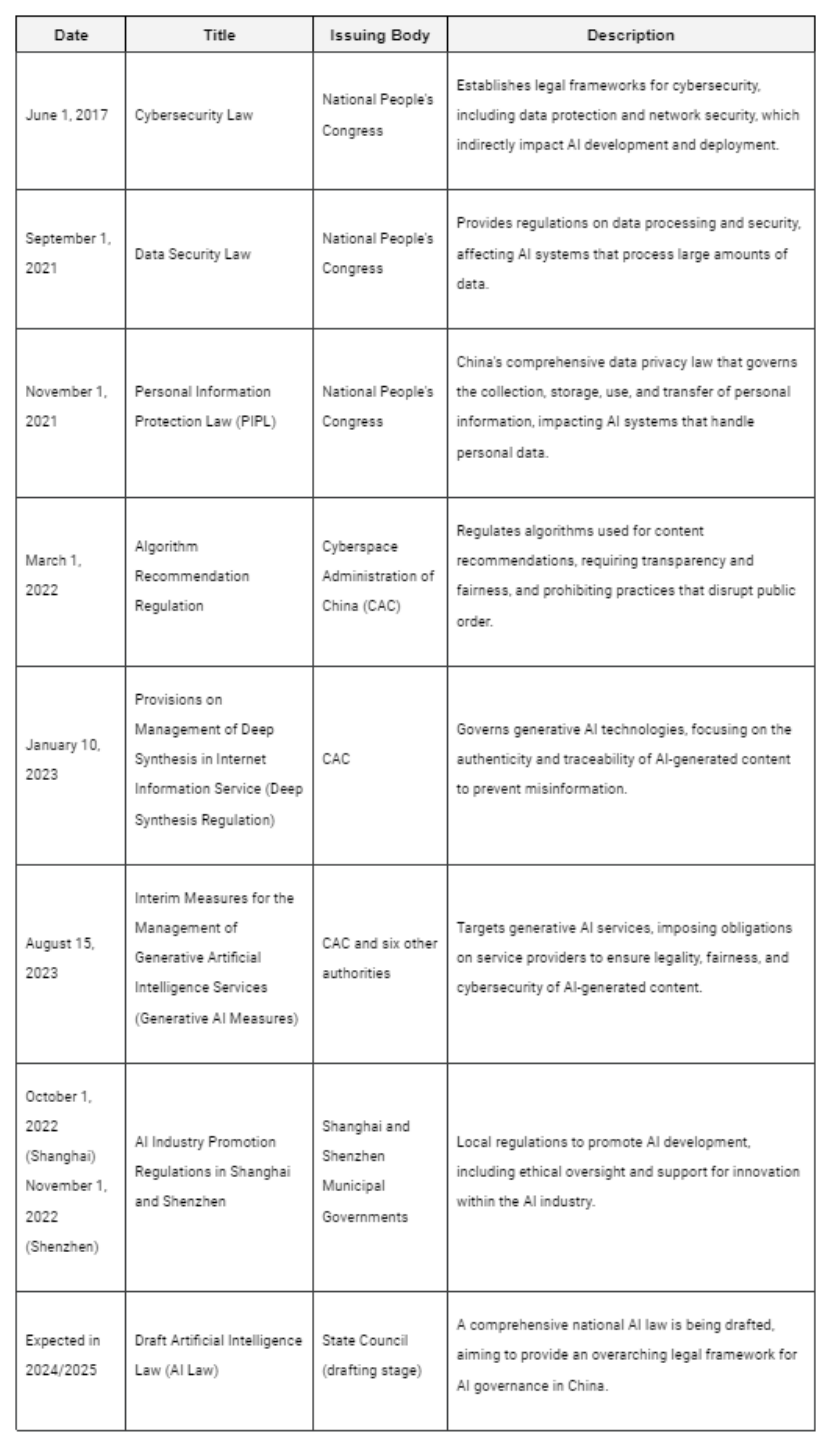} 
    \caption{Chinese AI Laws and Regulations}
    \label{fig:table_china_ai-laws-regulations.png}
\end{figure}

\FloatBarrier

\subsection{Registration}

\FloatBarrier

\begin{figure}
    \centering
    \includegraphics[width=0.95\linewidth]{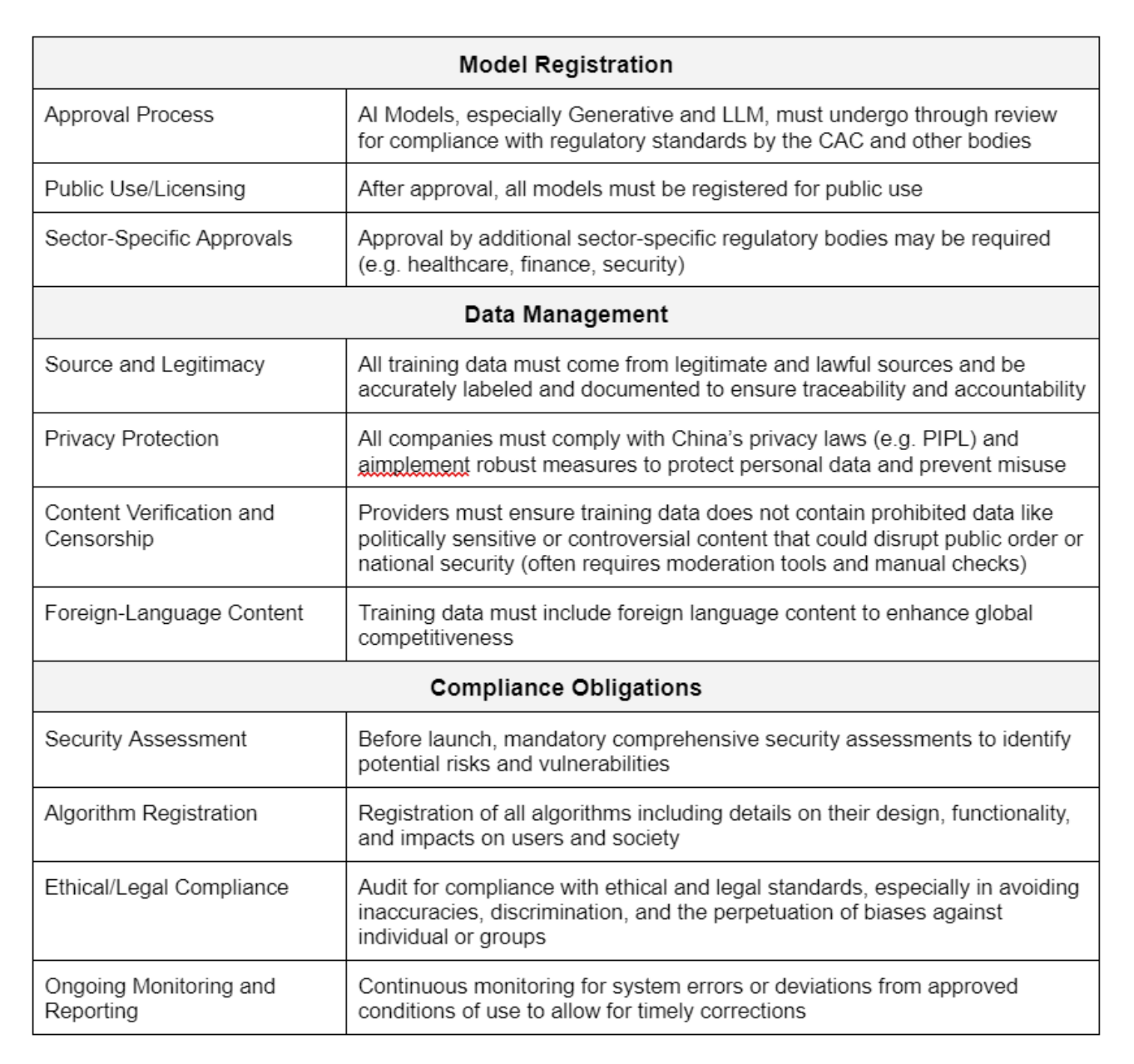} 
    \caption{ AI Model Compliance Steps in China}
    \label{fig:table_china_model_compliance}
\end{figure}

On paper, China has perhaps the most onerous AI regulation requirements of the three regions considered. Table 3 lists the three major steps for deploying advanced AI models (for example, Baidu’s LLM ERNIE) in order to be in compliance with regulatory laws (see Figure~\ref{fig:table_china_model_compliance}). These include model registration, rules for data management, and provisions for ongoing monitoring for compliance. The registration process alone illustrates how strict central regulation can slow down innovation and economic growth. As of March 2024, only 546 AI models have been registered, and just seventy are Large Language Models~\citep{chinamoney2024}. This number is in stark contrast to the countless commercial models, variants, and over 500,000 open-source LLMs on Huggingface.co~\citep{huggingface2024}, which is banned in China~\citep{chinatalk2023}.

\subsection{Compliance and Industrial Policy}

In 2015, China announced a national strategic plan and industrial policy called “Made In China 2025” or MIC2025 integrated with their 13th (2016-2020) and 14th (2021-2025) Five Year Plan~\citep{mic2015}. MIC2025 directs strong government support for innovation and high-end manufacturing to help make China a global leader in cutting-edge technologies like AI by 2025~\citep{crs2019}. Part of this plan calls for supporting 10,000 “Little Giants,” the small and mid-sized enterprises (SME) recognized as a key source of innovation~\citep{globaltimes2021}. Although large “National Champions'' like Baidu, Tencent, and Alibaba are expected to fully comply with AI regulations because of their dominant influence, the Little Giants are informally afforded leeway in order to avoid heavy regulatory burdens that could stifle innovation~\citep{zhang2024}. 

What this means from a practical standpoint is that despite such rigorous guidelines, enforcement in China is relatively lax. Startups and SMEs fly under the radar as long as they do not have a large public presence~\citep{zhang2022}. This approach allows for the promotion of  innovation, economic growth, and international competitiveness~\citep{yang2024}. 

China’s hybrid system of AI regulation and selective enforcement attempts to combine the strengths of both the EU and the US approaches. While regulatory guidance is generally light, top-level enforcement usually comes into play when destabilizing patterns arise. This reactive enforcement can cause transitory market disruptions and lead to strict and sometimes surprisingly punitive measures to reign in excesses and outcomes at odds with CCP values like “common prosperity”~\citep{caixin2021}. This pattern of regulatory crackdown is visible in other sectors from real estate\citep{bloomberg2021} to education~\citep{intresse2024}. Harsh penalties were levied by regulators between 2020-2022 to try to control excessive inequality and check the rise of powerful tech (Alibaba) and financial (Ant Group) corporations that could challenge government authority~\citep{chen2023}. Although deflating the real estate bubble significantly reduced household wealth tied to property speculation, the IMF shows China leads the world’s largest economies with a 5.2\% GDP growth~\citep{imf2024}. Some of this success is attributed to China’s strategic industrial policy with its flexible regulatory framework. 

\section{United States}

\subsection{Overview}

On October 30, 2023 US President Biden signed an executive order (EO \#14110) on the "Safe,Secure,and Trustworthy Development and Use of Artificial Intelligence"~\citep{whitehouse2023b}. This act moved beyond the voluntary commitments secured in July 2023~\citep{whitehouse2023a} and the October 2022 AI Bill of Rights~\citep{whitehouse2022}.  EO \#14110 represents the most comprehensive form of AI regulation in the United States to that date. It directly delegates AI responsibilities to over 50 existing federal regulatory agencies and other bodies with over 100 specific tasks designed to:

\begin{itemize}
  \item Build out the capacity to address emerging concerns around AI
  \item Integrate AI into agency operations
  \item Enhance coordination between agencies on AI-related matters
\end{itemize}

On August 28, 2024 the California Assembly passed SB 1047, the Safe and Secure Innovation for Frontier AI Models Act. Unlike the federal Presidental Executive Order, this state law focused on creating a regulatory framework to test, register, and audit models that could present a danger to public safety. This AI regulation targets models with substantial investment in pretraining and fine-tuning above given thresholds of \$100M/10\textsuperscript{26} flops and \$10M/10\textsuperscript{25} flops respectively.

\subsection{Laws and Regulations}

AI regulation in the US represents somewhat of a departure from the more typical US approach to regulation. In the US, the legislative branch typically passes laws that form the framework for regulation, which are then enforced by the executive branch, primarily under the oversight of various federal agencies. For example, the US Congress passes laws that define specific industries or activities along with broad goals such as advancing scientific research~\citep{nsf2024}, promoting fair markets~\citep{sec2024}, and safeguarding the environment~\citep{epa2024} (see NSF, SEC and EPA mission statements and goals). At times, multiple agencies will be tasked with regulating different aspects of the same broad goal. For example, the Federal Trade Commission (FTC), the Consumer Product Safety Commission (CPSC), and the Consumer Financial Protection Bureau (CFPB) specialize in different aspects of consumer protection and safety. US States and municipalities have also added regulations in areas they feel are inadequately addressed by federal regulations.

This approach is in keeping with historical norms. The philosophical distrust of centralized power is reflected in the very design of the American system of checks and balances between branches of government. A decentralized US approach is also a way to reduce bureaucratic layers, more directly empower domain experts, and balance power between competing narrow, self-interested parties. These include powerful voting blocks, special interests, and a \$46 billion state and federal lobbying industry~\citep{massoglia2024}. For these reasons, commercial applications of technology within the US have traditionally been regulated through various mechanisms including legislative action, executive orders, agency rulemaking, industry self-regulation, international agreements, and private self-regulation.

To remain competitive in rapidly changing world markets, US tech companies often pursue self-regulation as a strategy for tackling privacy, digital advertising, content moderation, and cybersecurity~\citep{cusumano2021, minow2023}. We see a similar approach taken in the Biden-Harris approach to securing voluntary commitments by leading AI companies to manage the risks posed by AI~\citep{whitehouse2023b}. Furthermore, international agreements or regulations are sometimes adopted by US companies to do business abroad, as in the case for EU’s GDPR~\citep{gdpr_article7} and China’s Cybersecurity Law~\citep{npc2016}. The regulatory process often combines these approaches, with laws providing the foundation for agency regulations involving public input and expert consultation as technologies and circumstances evolve.

The rapid pace of AI innovation and the immense potential impact of AI, coupled with the lack of technical expertise in government, has reversed the normal sequence for enacting regulation that begins with the U.S. Congress. EO \#14110 is a case where the executive branch is initiating many AI-related policies--from research to regulation--partly due to its ability to more quickly respond in a coherent and comprehensive manner \cite{whitehouse2023b}. Although somewhat exceptional for the US process of lawmaking, White House Presidential executive orders more closely match the top-down, centralized organization of the European Commission in Brussels and the CCP in Beijing. 

In spite of this similarity to the E.U. and China, aspects of the order nonetheless reflect the distinct US approach that can be characterized as “bottom-up” and distributed rather than “top-down” and centralized. In contrast to the more centralized, top-down approach to AI regulation prioritizing safety (EU) and social stability (China), the United States takes a more distributed, multi-stakeholder approach to AI regulation that mirrors its earlier approaches to regulating new technologies. 

While universal directives on AI are provided by the centralized political bodies of the CCP and to a lesser extent, the European Commission, a wide range of guidelines, initiatives, laws, and other policies including trade related to AI are distributed between various US federal branches and agencies and even states~\citep{perkins2024}. EO \#14110 organizes this distributed regulatory system with specific objectives and deadlines delegated to various federal agencies directly from the executive branch. 

Meanwhile, the US legislative branch is considering dozens of individual bills~\citep{govtrack2024}. Thune’s 2023 AI Research, Innovation and Accountability Act would create enforceable accountability and transparency for high-risk systems. The REAL Political Advertisements Act~\citep{klobuchar2023} aims to limit the use of Generative AI in campaigns, The Stop Spying Bosses Act~\citep{casey2023} aims to limit the use of AI by employers to surveil employees, and the No FAKES Act~\citep{coons2023} aims to protect visual and voice likenesses of individuals. Two notable, ambitious, and more restrictive plans have been introduced by Senator Schumer in the form of his SAFE initiative~\citep{schumer2023} and the Blumenthal-Hawley Framework~\citep{blumenthal2024}. At this time, however, none have been passed. Meanwhile, individual US states and municipalities have passed laws and are debating more extensive regulation regarding AI~\citep{iapp2024}. In 2023 more than 40 bills were proposed, and Texas and Connecticut adopted statutes focused on preventing discrimination~\citep{whitecase2024}. In the 2024 legislative session, at least forty states, Puerto Rico, The Virgin Islands and Washington D.C. have introduced AI bills and six states, Puerto Rico, the Virgin Islands have adopted resolutions or passed legislation~\citep{ncsl2023}. While Article VI of the US Constitution affirms the supremacy of federal law over state law, California is home to many of the largest AI corporations, and other industries (e.g. auto emission levels) have often complied with Calfornia's stricter standards.

\subsection{White House Executive Order 14110}

\FloatBarrier

\begin{figure}
    \centering
    \includegraphics[width=1.0\linewidth]{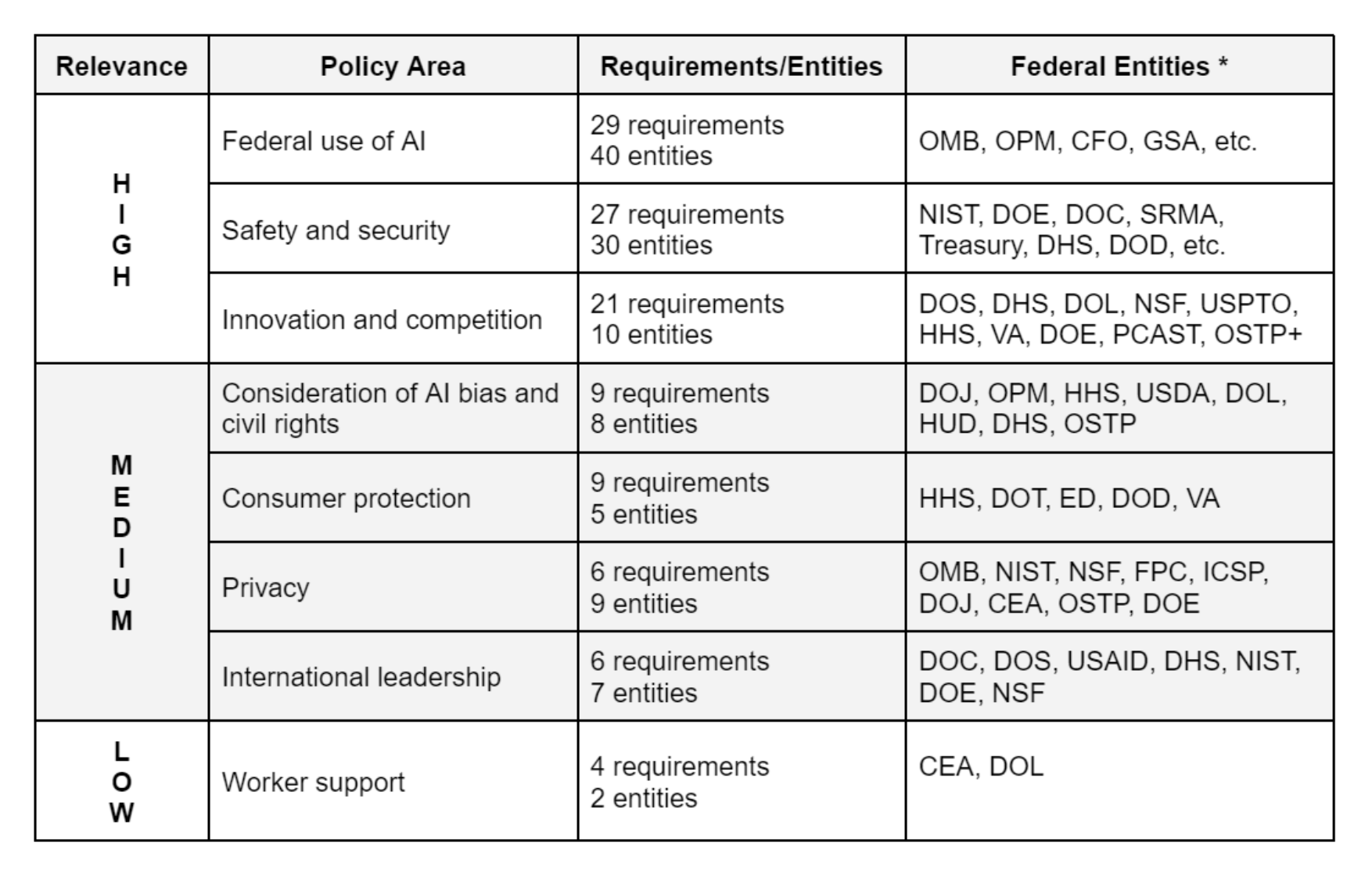} 
    \caption{Executive Order \#14110 on the Safe, Secure, and Trustworth Development and Use of AI (* see Appendix A for agency acronyms)}
    \label{fig:table_us_eo14110_agencies}
\end{figure}

Since 2016 and over three different Presidential administrations, a number of executive orders related to AI have been issued. The Biden White House’s October 2023 “Executive Order on the Safe, Secure, and Trustworthy Development and Use of Artificial Intelligence” is the most comprehensive to date~\citep{whitehouse2023b}. It directs over fifty federal agencies to take over one hundred specific actions addressing eight core policy areas listed in Figure \ref{fig:table_us_eo14110_agencies} including: safety and security, innovation and competition, worker support, bias and civil rights, consumer protection, privacy, federal use of AI, and international leadership~\citep{whitehouse2024a}. The eight policy areas are ranked by the aggregate number of requirements and federal entities assigned to each area. These arguably provide a loose sense of priorities in each policy area from the most relevant (federal use, safety/security, and innovation/competition) to the least (worker support). EO \#14110 implements many guidelines in the 2022 AI Bill of Rights to ensure the responsible design and use of artificial intelligence with regards to civil rights and privacy in areas such as hiring, healthcare, and surveillance~\citep{whitehouse2022}.

EO \#14110 also addresses many of the core concerns highlighted in the EU AI Act. It does so, however, with several key differences~\citep{crs2024}. While the EU AI Act establishes a new regulatory agency, the EU AI Office, which coordinates with member states, industry and civil society, the current US strategy relies upon augmenting the extensive network of existing US federal agencies with pre-existing specialized domain expertise. The US approach can be seen to emphasize extending expansive regulatory and legal frameworks from the ground up where infrastructure already exists, in contrast to creating a new centralized regulatory framework.

Because the US approach involves over fifty federal agencies, it is also much more extensive in implementation details than the EU. It directly addresses broader issues like unemployment, education, research, and consumer protection. Finally, and again in contrast to the EU AI Act, this US strategy is arguably more immediately actionable given the over one hundred specific objectives. Many deadlines are delegated to federal agencies to be completed within 180 to 270 days. These agencies are already specialized across a broad spectrum of existing federal government responsibilities that are being disrupted by AI.  As of this writing, both the White House 180-day and 270-day deadlines have been met~\citep{whitehouse2024b, whitehouse2024d}. 

Enforcement is another major area of difference between the EU and the US. The EU AI Acts’ risk model is premised upon prevention. General guidelines, specific penalties, and centralized regulation prohibit activities unless explicitly permitted. In contrast, the current US risk model is permissive. It promotes innovation through competition, encourages decentralized self-regulation, and relies upon an extensive network of existing laws and regulations against abusive, illegal, and negligent practices. These networks of existing laws range over a wide spectrum, extending from specific consumer protection laws to evolving intellectual property laws~\citep{walters2022}. This permissive approach follows the American tradition of tech sector self-regulation with its notable success in sectors like online advertising (DAA, NAI), cybersecurity (NIST, CISA), biotechnology (IGSC, IASB), nanotechnology (ISO, NanoRisk), and cloud computing (CSA). 

Within this permissive structure, and in response to the White House EO 14110, the National Institute of Standards and Technology~\citep{nist2024a} established the US AI Safety Institute in January of 2024. Housed within the larger Department of Commerce, NIST was originally established to facilitate U.S. industrial competitiveness. The US AI Safety Institute has members from academia, industry, and nonprofits. This partnership mirrors the kind of decentralized and voluntary approach discussed earlier. The US AI Safety Institute’s initial task forces focus on safety, evaluation, measurement, and risk management. Their work follows upon the initial Risk Management Framework published in April 2024~\citep{nist2024b}. On July 12 2024 representatives from the US AI Safety Institute and the European AI Office met in Washington, D.C. and announced plans for further cooperation and collaboration~\citep{nist2024c}.

\FloatBarrier

\subsection{California SB 1047}

\FloatBarrier

\begin{figure}
    \centering
    \includegraphics[width=0.9\linewidth]{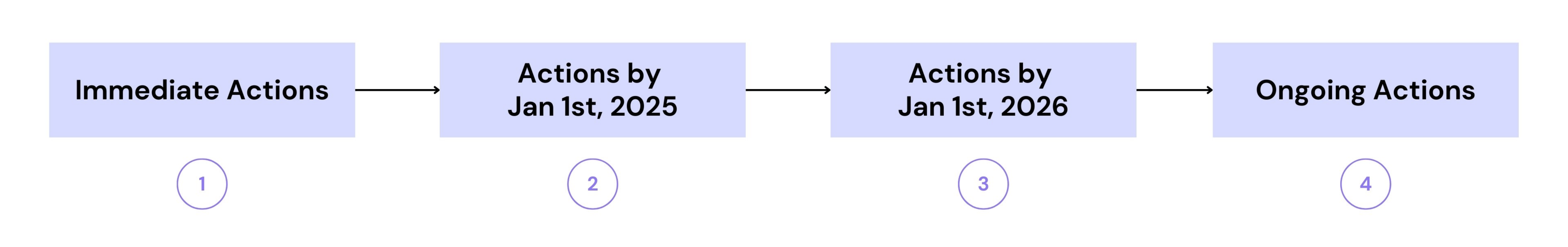} 
    \caption{Action Timeline for SB 1047}
    \label{fig:fig_sb1047_timeline}
\end{figure}

The California Senate Bill 1047 (SB 1047: Safe and Secure Innovation for Frontier AI Models Act), introduced in February 2024 by Senator Scott Wiener, attempts to minimize potential negative societal impacts of AI in the face of rapid progress~\citep{LegiScan2024}. The bill emerged in response to growing concerns about the unique threats posed by powerful AI systems~\citep{gdprlocal2024}. It would affect nearly every leading AI company either based in Silicon Valley or doing business with California, the 5th largest economy in the world.

SB 1047 aims to establish a comprehensive AI regulatory framework in California that is more narrowly focused on 'frontier models' as defined by potential risks and computational resources~\citep{ktslaw2024}. The bill has garnered support from a diverse coalition of politicians and stakeholders including AI researchers, unions, and some technology companies advocating for responsible AI development~\citep{thehill_ai_supporters_2024, lovely2024, verge2024}. It has also faced opposition from politicians, industry groups, some AI researchers, and open-source advocates who worry about overly stringent regulations. Critics fear the bill could stifle innovation, favor a handful of tech giants, and put California at a competitive disadvantage~\citep{nunez2024, abbott2024coalition_2024, chilson_coalition_letter_2024, abboud_eus_2023}. However, opinions are roughly split within traditional interest groups including large corporations, leading researchers, and politicians across both parties, leaving the future of SB 1047 somewhat uncertain.

Ongoing debates surrounding the bill's potential impact on US competitiveness, AI research and development continue. Uncertainties around costs, definition of key terms, and the feasibility of implementing requirements raise more immediate concerns. Much debate centers on the balance between fostering innovation and implementing safeguards~\citep{foley2024, morganlewis2024}. One of the most controversial aspects of SB 1047 is its requirement for developers to implement a full shutdown capability for covered AI models~\citep{neontri2024}. Concerns have been raised over potential disruptions to critical infrastructure and services that may rely on these systems, a demonstration of the complex interdependencies and deep integration of AI with various sectors of the economy and society~\citep{LegiScan2024}.

Since its introduction, SB 1047 has undergone several revisions in response to feedback from various stakeholders. Notable changes include narrowing the scope of pre-harm enforcement and potential liabilities based on site-specific plans. Controversial elements like the frontier model division, know your customer requirements, and uniform pricing were removed.  Ambiguities in definitions like "covered models" were clarified to more precisely target high-impact AI systems. The scope of the required safety and security protocols were defined, and the timeline for compliance was adjusted to give companies more time to adapt to the new regulations~\citep{LegiScan2024}. These amendments demonstrate the iterative nature of the legislative process, particularly when dealing with rapidly-evolving technologies like AI and the conflicting interests of diverse stakeholders. As other jurisdictions both within the United States and internationally grapple with similar issues, the outcome of California's legislative efforts may have far-reaching impacts for the future of AI regulation and development~\citep{foley2024}.

Unlike the EU AI Act, which adopts a comprehensive, risk-based approach to AI regulation across various sectors and applications, SB 1047 appears to focus more narrowly on high-impact AI systems, particularly those trained using substantial computational resources~\citep{cmswire2024, euronews2024}. This targeted approach may reflect a philosophy that prioritizes regulating the most powerful and potentially influential AI models that have the greatest far-reaching societal impacts.

While the EU AI Act offers a degree of flexibility in implementation, allowing for tailored requirements for specific high-risk AI applications~\citep{uchicagobusinesslaw2024}, it is unclear whether SB 1047 adopts a similar approach.  In contrast, the federal U.S. approach to AI regulation, which may influence SB 1047, typically involves adapting existing laws and regulatory frameworks to address AI-specific challenges. The special focus on the computational resources used to train frontier AI models suggest a unique regulatory philosophy in which the technical potential of AI systems emerges as the key factor in determining regulatory requirements.

As shown in Figure \ref{fig:fig_sb1047_timeline}, SB 1047's timeline for implementation and compliance, the bill is structured around specific milestones: immediate actions, actions required by January 1, 2026, actions required by January 1, 2027, and ongoing actions~\citep{smdailyjournal2024}. While detailed information about the specific requirements associated with each time frame is not available, this phased approach suggests a recognition of the need for a gradual implementation process, allowing stakeholders time to adapt to new regulatory requirements.

\FloatBarrier

\subsubsection{Immediate Actions}

\begin{figure}
    \centering
    \includegraphics[width=0.7\linewidth]{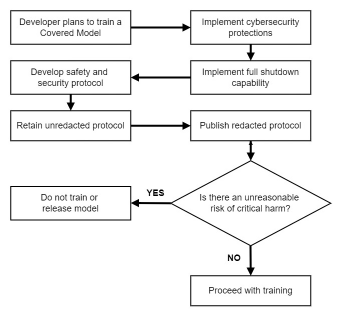} 
    \caption{Immediate actions required according to CA SB 1047}
    \label{fig:fig_sb1047_immediate}
\end{figure}

Figure \ref{fig:fig_sb1047_immediate} outlines a series of immediate actions that developers of covered AI models must undertake if SB 1047 is signed into law. These actions are designed to establish a framework for the responsible development, risk management, and public safety around the deployment of high-impact AI systems.

At the core of SB 1047's immediate requirements is the establishment of administrative, technical, and physical measures to prevent unauthorized access and misuse of covered models, with particular focus on defending against advanced persistent threats and sophisticated actors. However, the bill's specific cybersecurity requirements are not explicitly detailed.

Another key immediate action is the development of a safety and security protocol. This written document outlines procedures for managing risks throughout the model's lifecycle, including testing procedures to assess potential harm. Again, exact details of this protocol are not provided.

\FloatBarrier

\subsubsection{Actions by January 2026}

\begin{figure}
    \centering
    \includegraphics[width=\linewidth]{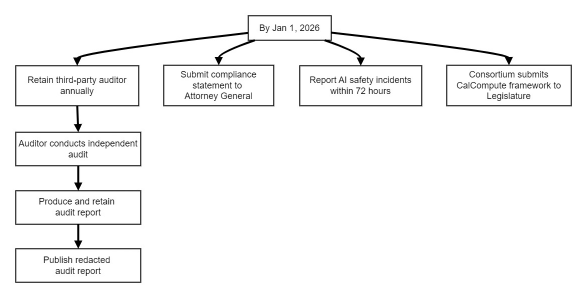}
    \caption{Actions required by Jan 1st, 2026 according to CA SB 1047}
    \label{fig:fig_sb1047_jan2026}
\end{figure}

SB 1047 outlines a series of actions that developers of covered AI models must undertake by January 1, 2026, as illustrated in Figure \ref{fig:fig_sb1047_jan2026}. A key requirement is the implementation of annual third-party audits. Starting January 1, 2026, developers must engage independent auditors to assess their compliance with the bill's requirements. These audits, conducted according to regulations issued by the Government Operations Agency, require detailed reports evaluating internal controls and any instances of noncompliance. Developers are required to retain unredacted versions of these audit reports and provide access to the Attorney General upon request. In line with the bill's commitment to transparency, redacted versions of these reports must be published, with redactions limited to protecting sensitive information.

SB 1047 also mandates that developers submit annual compliance statements to the Attorney General. These statements, signed by the chief technology officer or a senior corporate officer, must include assessments of potential critical harms and verification of compliance with the bill's requirements. The bill also introduces a 72-hour reporting requirement for any AI safety incidents affecting covered models.

Furthermore, SB 1047 establishes a consortium within the Government Operations Agency tasked with developing a framework for "CalCompute," a public cloud computing cluster. This initiative, culminating in a report to be submitted to the Legislature by January 1, 2026, demonstrates a commitment to creating public infrastructure for AI development and research. These actions, taken together, demonstrate the desire to create a more robust, transparent, and accountable ecosystem for AI development and deployment.

\FloatBarrier

\subsubsection{Actions by January 2027} 

\begin{figure}
    \centering
    \includegraphics[width=\linewidth]{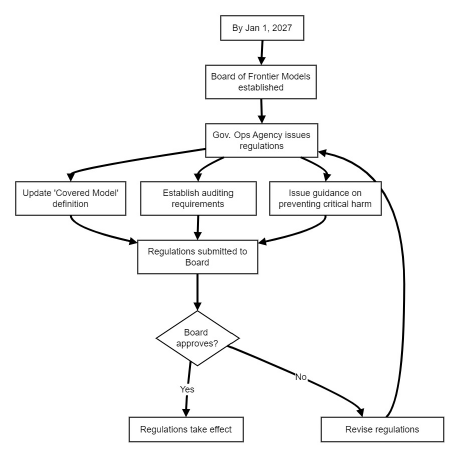}
    \caption{Actions required by Jan 1st, 2027 according to CA SB 1047}
    \label{fig:fig_sb1047_jan2027}
\end{figure}

SB 1047 also outlines a series of actions to be implemented by January 1, 2027, as illustrated in Figure \ref{fig:fig_sb1047_jan2027}. These actions are designed to establish a robust regulatory framework and governance structure for high-impact AI systems in California.

Actions include the establishment of the Board of Frontier Models within the Government Operations Agency. As shown in Figure \ref{fig:fig_sb1047_jan2027}, the board will consist of nine members with expertise in AI, safety, and related fields. Their primary responsibility will be to approve regulations and guidance, ensuring that the regulatory framework remains informed by the latest developments in AI technology and safety considerations.

By January 1, 2027, and annually thereafter, the Government Operations Agency is mandated to issue a set of regulations subject to approval by the Board of Frontier Models. These regulations serve three functions. First, they update the definition of "covered model," adjusting thresholds to reflect technological advancements and emerging risks. Second, they establish comprehensive auditing requirements, defining standards and best practices for the third-party audits introduced in the previous phase. Finally, they provide guidance on preventing critical harm and offer recommendations to developers on risk mitigation strategies.

\FloatBarrier

\subsubsection{Ongoing Actions}

\begin{figure}
    \centering
    \includegraphics[width=\linewidth]{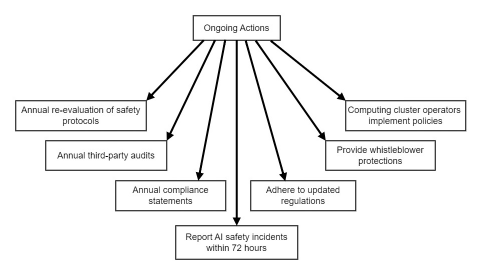}
    \caption{Ongoing Actions required according to CA SB 1047}
    \label{fig:fig_sb1047_ongoing}
\end{figure}

Finally, SB 1047 outlines a comprehensive set of ongoing actions that developers and operators of high-impact AI systems must undertake to ensure continued compliance and safety, as illustrated in Figure \ref{fig:fig_sb1047_ongoing}. First and foremost, developers are required to engage in annual re-evaluations and updates of their safety and security protocols. This process ensures that protocols remain current and responsive to changes in model capabilities and industry best practices. Complementing this internal review, the bill mandates the continuation of annual third-party audits.

\sloppy
Figure \ref{fig:fig_sb1047_ongoing} highlights the ongoing obligation for developers to report AI safety incidents within 72 hours of discovery. The bill also requires adherence to updated regulations issued by the Government Operations Agency: developers must remain compliant with the latest standards and definitions.

Of particular note is that SB 1047 introduces important protections for whistleblowers, prohibiting retaliation against employees who report non-compliance or risks. The bill also outlines specific responsibilities for operators of computing clusters. These include policies to assess customers' use of compute resources with a focus on those using resources sufficient to train covered models, along with a capability to enact full shutdowns if necessary.

\FloatBarrier

\subsubsection{Enforcement}

SB 1047 establishes an enforcement mechanism to ensure compliance with its provisions. It is designed to provide the necessary authority and tools to address violations and protect the public interest. The Attorney General is granted significant authority in enforcing SB 1047 and is empowered to bring civil actions against entities that violate the bill's provisions. This authority extends to halting non-compliant activities and seeking a range of remedies, including civil penalties and damages where appropriate. 

The penalty structure is designed to be proportionate and impactful. Civil penalties are calculated based on the cost of compute used to train the model in question. This approach ensures that penalties are commensurate with the scale and potential impact of the AI system involved. Additionally, the bill provides for specific penalties for particular violations such as misrepresentation by auditors.

Protections and remedies are also established for whistleblowers. Employees who face retaliation for reporting non-compliance are granted the right to seek injunctive relief. These protections are cumulative with other laws, ensuring safeguards for individuals who come forward with concerns. 

\subsubsection{Veto and After Effects}

On September 29th 2024, Governor Newsom vetoed SB 1047 \citep{newsom2024veto} because he opposed standards solely based on model size and computational resources. Instead, he said it was necessary to consider whether AI systems would be deployed in high-risk environments and involve critical decision-making. As an alternative, Newsom will consult with  AI experts to develop more targeted guardrails for AI deployment and work with the legislature on more empirically-based regulation. Despite the veto, SB 1047 has significantly impacted the AI regulation debate by:

\begin{itemize}
    \item Highlighting the need for proactive safety measures and accountability in AI development.
    \item Sparking discussions on the appropriate metrics for regulating AI, such as computational resources versus actual risks and impacts.
    \item Raising awareness about potential catastrophic risks associated with advanced AI systems.
    \item Demonstrating the challenges of balancing innovation with safety concerns.
    \item Encouraging industry stakeholders, startups, investors, and open-source developers to seriously consider AI safety and governance in their strategic planning.
\end{itemize}

The bill and subsequent veto have also revealed tensions between state-level and national approaches to AI regulation, as well as unlikely divisions and alliances within academia, industry, and government (see Appendix A). Many argue that a federal framework would be the more effective place to address AI regulation. Nonetheless, the debate over California SB 1047 will undoubtedly influence future regulatory efforts at both state and federal levels and could lead to more nuanced and effective AI regulation. Indeed, Governor Newsom has signed 17 AI-related bills in the month before his SB 1047 veto and established new initiatives with the CA legislature and AI experts to establish workable guardrails and empirical, science-based trajectory analysis of frontier models \citep{newsom2024postveto}.

The bill and subsequent veto have also revealed tensions between state-level and national approaches to AI regulation, as well as unlikely divisions and alliances within academia, industry, and government (see Appendix A). Many argue that a federal framework would be more effective to address AI regulation. Nonetheless, the debate over California SB 1047 will likely influence future regulatory efforts at both state and federal levels. Governor Newsom signed an impressive 17 AI-related bills prior to his SB 1047 veto. Immediately after his veto, Newsom launched initiatives with the legislature and AI experts to develop workable AI guardrails. These initiatives also aim to create empirically-based predictions of future frontier model capabilities likely to target specific high-risk AI applications.

\section{Conclusion}

The EU, China, and the US are each evolving distinct regulatory systems that vary in approach and emphasis. The EU AI Act proposes a coherent, universal, risk-based regulatory framework with strict and well-defined penalties. It is criticized, however, for stifling innovation, using ambiguous language, and not anticipating expected challenges in implementation across heterogeneous use cases. The Chinese approach to AI regulation synthesizes the US approach of use-case specific laws with general guidelines translated into a centralized and comprehensive registration, testing, and monitoring framework. At the same time, innovation and economic growth is directly and indirectly supported by initiatives like investment in thousands of ‘Little Dragons’ alongside relatively lax enforcement for SMEs. This hybrid approach has led to a variety of surprising technological breakthroughs and economic successes, but it risks criticism of capriciousness given that regulations are not uniformly applied or enforced.  The Biden White House Executive Order on “Safe, Secure, and Trustworthy Development and Use of Artificial Intelligence” is the most organized plan in the US. It delegates over one hundred specific tasks to over fifty federal agencies in order to build out AI expertise and oversight according to specific domain expertise. The decentralized US approach also involves smaller regulatory initiatives by the US Congress, individual states like California, and even cities. This reflects the US market-driven approach acknowledging competing stakeholders. The approach has come under criticism, however, for relying too heavily on self-regulation and for being susceptible to flaws like regulatory capture. In response, California, home to most of the leading AI companies, introduced what is arguably the most stringent regulation. After ten major revisions, Governor Newsom vetoed the bill. 

Growing trade tensions and geopolitical competition between the US and China are bolstering arguments for AI regulation policies favoring faster innovation and technological independence aligned with industrial policy. The US, with China following suite, has increased tariffs, coordinated international export bans, and imposed sanctions on strategic technologies like EVs, advanced chips, and semiconductor manufacturing equipment. These geopolitical tensions mean that the regulatory landscape will continue to evolve as countries re-evaluate their stance towards risk alongside their desire to remain at the forefront of AI development.

\section{Contributions}
\begin{itemize}
  \item Jon Chun: The United States and China
  \item Christian Schroeder de Witt: The European Union
  \item Katherine Elkins: The United States
\end{itemize}
Each author has reviewed each other's section and takes responsibility for the content of their own.


\FloatBarrier  

\bibliographystyle{abbrvnat}
\bibliography{main}

\clearpage  

\section*{Appendix A: Sample of Diverse Supporters and Opponents of California SB 1047}
\addcontentsline{toc}{section}{Appendix A: Sample of Diverse Supporters and Opponents of California SB 1047}

\small
\begin{longtable}{|p{0.12\textwidth}|p{0.18\textwidth}|p{0.20\textwidth}|p{0.40\textwidth}|}
\hline
\textbf{Side} & \textbf{Category} & \textbf{Name} & \textbf{Description} \\ \hline
Pro & Politician & Sen. Anthony Wiener & \url{https://tinyurl.com/4ub8u7ha} \\ \hline
Pro & Academics & Yoshua Bengio, Geoffrey Hinton, Lawrence Lessig, Stuart Russell & \url{https://safesecureai.org/experts} \\ \hline
Pro & Academic & Dan Hendrycks & \url{https://tinyurl.com/ytn7vcfr} \\ \hline
Pro & AI Researchers & Call to Lead & \url{https://calltolead.org/} \\ \hline
Pro & AI Non-Profit & Center for AI Safety & \url{https://www.safe.ai/work/statement-on-ai-risk} \\ \hline
Pro & AI Frontier Corp & OpenAI & \url{https://tinyurl.com/2s42ndr5} \\ \hline
Pro & AI Frontier Corp & Anthropic & \url{https://tinyurl.com/yth2k2hn} \\ \hline
Pro & Hollywood & Actors and Unions & \url{https://tinyurl.com/4ad85wzu} \\ \hline
Pro & Entrepreneur & Elon Musk & \url{https://x.com/elonmusk/status/1828205685386936567} \\ \hline
Con & Politician & Sen. Nancy Pelosi & \url{https://tinyurl.com/y25yrre9} \\ \hline
Con & Politician & Rep. Ro Khanna & \url{https://tinyurl.com/3sz3jwns} \\ \hline
Con & Politician & SF Mayor London Breed & \url{https://tinyurl.com/3664wwua} \\ \hline
Con & Academic & Yan LeCun & \url{https://x.com/ylecun/status/1807552057466909156} \\ \hline
Con & Academic & Fei-Fei Li & \url{https://tinyurl.com/4b9uzctb} \\ \hline
Con & Academics & CalTech & \url{https://tinyurl.com/2xf6xcw5} \\ \hline
Con & AI Frontier Corp & Meta & \url{https://tinyurl.com/y2sus9xp} \\ \hline
Con & AI Frontier Corp & Google & \url{https://tinyurl.com/adseybwp} \\ \hline
Con & AI Non-Profit & The AI Alliance & \url{https://thealliance.ai/core-projects/sb1047} \\ \hline
Con & Thought Leaders & Investors, Entrepeneurs and Academics & \url{https://tinyurl.com/26h9f397} \\ \hline
\caption{Split among leading voices on California SB 1047 does not follow traditional dividing lines}
\end{longtable}

\end{document}